\documentclass[12pt]{article}
\usepackage{a4wide}
\usepackage{amssymb}
\usepackage{amsmath}
\usepackage{graphicx}
\usepackage{slashed}
\usepackage{verbatim}

\def\XXint#1#2#3{{\setbox0=\hbox{$#1{#2#3}{\int}$}
     \vcenter{\hbox{$#2#3$}}\kern-.52\wd0}}

\newcommand{\be}{\begin{equation}}\newcommand{\ee}{\end{equation}}
\newcommand{\bea}{\begin{eqnarray}} \newcommand{\eea}{\end{eqnarray}}

\def\makeatletter{\catcode`\@=11}
\makeatletter
\def\mathbox#1{\hbox{$\m@th#1$}}%
\def\math@ccstyles#1#2#3#4#5#6#7{{\leavevmode
      \setbox0\mathbox{#6#7}%
      \setbox2\mathbox{#4#5}%
      \dimen@ #3%
      \baselineskip\z@\lineskiplimit#1\lineskip\z@
      \vbox{\ialign{##\crcr
             \hfil \kern #2\box2 \hfil\crcr
             \noalign{\kern\dimen@}%
             \hfil\box0\hfil\crcr}}}}
\def\mathaccstyles{\math@ccstyles\maxdimen}
\def\maththroughstyles{\math@ccstyles{-\maxdimen}}
\def\unity%
 {\maththroughstyles{.45\ht0}\z@\displaystyle {\mathchar"006C}\displaystyle 1}

\makeatletter \@addtoreset{equation}{section} \makeatother

\begin{document}

\setcounter{table}{0}

\begin{flushright}\footnotesize

\texttt{ICCUB-17-020}
\vspace{0.6cm}
\end{flushright}

\mbox{}
\vspace{0truecm}
\linespread{1.1}

\centerline{\LARGE \bf Boundary Conformal Anomalies}
\medskip

\centerline{\LARGE \bf on Hyperbolic Spaces and Euclidean Balls}

\vspace{.5cm}

 \centerline{\LARGE \bf }

\vspace{1.5truecm}

\centerline{
    {\large \bf Diego Rodriguez-Gomez${}^{a}$} \footnote{d.rodriguez.gomez@uniovi.es}
    {\bf and}
    {\large \bf Jorge G. Russo ${}^{b,c}$} \footnote{jorge.russo@icrea.cat}}

\vspace{1cm}
\centerline{{\it ${}^a$ Department of Physics, Universidad de Oviedo}} \centerline{{\it Avda.~Calvo Sotelo 18, 33007  Oviedo, Spain}}
\medskip
\centerline{{\it ${}^b$ Instituci\'o Catalana de Recerca i Estudis Avan\c{c}ats (ICREA)}} \centerline{{\it Pg.Lluis Companys, 23, 08010 Barcelona, Spain}}
\medskip
\centerline{{\it ${}^c$ Departament de F\' \i sica Cu\' antica i Astrof\'\i sica and Institut de Ci\`encies del Cosmos}} \centerline{{\it Universitat de Barcelona, Mart\'i Franqu\`es, 1, 08028
Barcelona, Spain }}
\vspace{1cm}

\centerline{\bf ABSTRACT}
\medskip

\noindent 

We compute conformal anomalies for conformal field theories with free conformal scalars and massless spin $1/2$ fields in hyperbolic space $\mathbb{H}^d$ and in the ball $\mathbb{B}^d$,
for $2\leq d\leq 7$. These spaces are related by a conformal transformation.
In even dimensional spaces, the conformal anomalies on $\mathbb{H}^{2n}$ and $\mathbb{B}^{2n}$
are shown to be identical. In odd dimensional spaces, the conformal anomaly on $\mathbb{B}^{2n+1}$ comes from a boundary contribution, which exactly coincides with that of
$\mathbb{H}^{2n+1}$  provided one identifies the UV short-distance cutoff on   $\mathbb{B}^{2n+1}$ with the inverse large distance IR cutoff on $\mathbb{H}^{2n+1}$,
just as prescribed by the conformal map. 
As an application, we determine, for the first time, the conformal anomaly coefficients
multiplying the Euler characteristic of the boundary for scalars and half-spin fields with various boundary conditions in $d=5$ and $d=7$.

\newpage

\tableofcontents


\section{Introduction and conclusions}

Recently, the study of conformal field theories on manifolds with a boundary has received
increasing attention \footnote{For early discussions, see {\it e.g.}  \cite{Dowker:1989ue}.}  \cite{Fursaev:2013mxa}--\cite{Herzog:2017kkj}.
These theories are of great interest because of their many applications in condensed matter, quantum field theory and string theory.
Understanding the complete structure of conformal anomalies due to boundary contributions is an important open problem.

A particularly interesting class of spaces with boundaries has been recently introduced in \cite{Rodriguez-Gomez:2017kxf}, where $d$ dimensional conformally flat spaces of the form $\mathbb{S}^a\times \mathbb{H}^{d-a}$ have been considered. These spaces generalize the celebrated $\mathbb{S}^1\times \mathbb{H}^{d-1}$ space, whose mapping to $\mathbb{R}^d$ covers only the causal development of an $\mathbb{S}^{d-1}$ and hence allows for the computation of the entanglement entropy across the sphere in terms of a standard thermal partition function. 

Among the spaces introduced in \cite{Rodriguez-Gomez:2017kxf}, an
interesting class is the case of odd $d$ and even $a$, i.e. the direct product of an even-dimensional sphere and an odd-dimensional hyperbolic space. On general grounds, because the space is odd-dimensional, one would not expect a bulk conformal anomaly. Nevertheless, as shown in \cite{Rodriguez-Gomez:2017kxf}, a conformal anomaly arises due to the fact that the boundary of the odd-dimensional hyperbolic space is an even-dimensional sphere which allows for an anomaly term proportional to its Euler number.
Technically, the anomaly arises  through the regularization of the divergent volume of the odd-dimensional hyperbolic space, which may lead to the somewhat surprising conclusion that the anomaly is an IR effect. Note that in fact the same puzzle arises in the $\mathbb{S}^1\times \mathbb{H}^{d-1}$ case for even $d$.

In this paper we will clarify this issue by considering the particular case of the $\mathbb{H}^d$ spaces. These can be conformally mapped into the ball $\mathbb{B}^d$ with no topology changing, so that the partition functions, as we show below, agree. However, in the $\mathbb{B}^d$ case, the divergence leading to the anomaly is due to short distance (to the boundary) effects, and it is therefore a UV effect. 
This shows that also in the $\mathbb{H}^d$ case the boundary anomalies can be regarded  as a UV effect, both for even and odd $d$.

The fact that the naively-looking IR regulator in $\mathbb{H}^d$ is secretly a UV regulator in the ball $\mathbb{B}^d$
can be directly understood considering the mapping among the two spaces. To be explicit, the $\mathbb{H}^d$ space of radius $R$ can be conformally mapped into $\mathbb{B}^d$ of the same radius as

\bea
\label{metrics}
ds^2 &=& R^2\,\Big[dy^2+\sinh^2y\,d\Omega_{d-1}^2\Big]=
\frac{d\rho^2}{1+\frac{\rho^2}{R^2}}+\rho^2 d\Omega_{d-1}^2
\nonumber\\
&=& \Big( \frac{2}{1-\frac{r^2}{R^2}}\Big)^2\,\big[ dr^2+r^2d\Omega_{d-1}^2\big]\, .
\label{muno}
\eea
In particular, the dictionary between radial coordinates is $\tanh \frac{y}{2} = \frac{r}{R}$ and
\begin{equation}
\label{map}
\rho = \frac{2r}{1-\frac{r^2}{R^2}} \, .
\end{equation}
Note that from here it follows that $r\in[0,R]$.\footnote{One may also consider introducing a dimensionless coordinate $\varrho=\frac{r}{R}\in[0,1]$, so that the $\mathbb{B}^n$ metric becomes $ds^2=R^2\Big[d\varrho^2+\varrho^2\,d\Omega_{d-1}^2\Big].$} Moreover, from \eqref{map} it also follows that the boundary of the $\mathbb{H}^d$ at large $\rho$ is mapped to the boundary of the $\mathbb{B}^d$ at $r\sim R$. 
An IR cutoff at $\rho=L$ corresponds to a UV cutoff at $R-r=\delta$, with
\be
\label{cutoffrel}
L=\frac{R^2}{\delta}\ .
\ee
%
%
Thus $L$ is to be interpreted as the large volume (IR) cut-off for the $\mathbb{H}^d$ while $\delta = R-r$ is to be interpreted as a short distance (UV) cut-off for the $\mathbb{B}^d$ (in fact, we will use the same notation $\delta $ for UV short-distance cutoffs in all spaces). Then, \eqref{cutoffrel} implements explicitly the relation between the IR regulator in the $\mathbb{H}^d$ and the UV regulator for the $\mathbb{B}^d$. It is interesting to note that this UV/IR mixing is very reminiscent to the $AdS/CFT$ correspondence, and may suggest that a sort of ``rigid holography" is at play.

A second reason to consider spaces of the form $\mathbb{H}^d$ or $\mathbb{B}^d$ is that these spaces have as boundary $\mathbb{S}^{d-1}$. As discussed below,  one would expect that the only possible (boundary) anomaly term which can be constructed is proportional to the Euler number of the sphere. Indeed, even though the full structure of boundary anomalies is not known beyond $d=4$ \cite{Herzog:2015ioa,Solodukhin:2015eca,Fursaev:2016inw}, 
one can separate contributions to the boundary anomaly proportional to the Euler number of the boundary (``$A$-type" anomaly)
from contributions constructed in terms of the Weyl tensor and the traceless part of the extrinsic curvature of the boundary \cite{Solodukhin:2015eca}.
  {}From our computation, we can read off the boundary $a$ central charges for scalars and spin $1/2$ fields in different dimensions. We stress that the new anomaly coefficients found here (multiplying the Euler characteristic of the boundary) should be universal and apply to any conformal field theory on a manifold with boundary (i.e. not only to $\mathbb{H}^{2n+1}$ or  $\mathbb{B}^{2n+1}$).

The remainder of the paper is structured as follows. In section \ref{setting} we briefly describe a standard method to compute the 
conformal anomaly that will be used here. In section \ref{even} we compute the logarithmic part of the free energy for scalars and fermions on even-dimensional $\mathbb{H}^{2n}$ and $\mathbb{B}^{2n}$ spaces, and find that they match. In section \ref{odd} we turn to the odd dimensional spaces $\mathbb{H}^{2n+1}$ and $\mathbb{B}^{2n+1}$. Like in the even dimensional case, we find that the logarithmic terms in the corresponding free energies match. In section \ref{centralcharges} we use these results to determine the values of the $a$-boundary central charges for scalars and fermions in different dimensions, for the different possible boundary conditions. 

\section{Setting the computation}\label{setting}

In the following we will be interested on conformal anomalies either on $\mathbb{H}^n$ or on $\mathbb{B}^n$. These show up on the free energy, since it may contain a logarithmic term of the form\footnote{We will denote bosonic free energies and $A$-anomaly coefficients on $\mathbb{H}^d$ or $\mathbb{B}^d$ as $F^{\mathbb{H}/\mathbb{B}}_d$ and $A^{\mathbb{H}/\mathbb{B}}_d$ respectively. In turn, we will reserve calligraphic fonts for their fermion counterparts,  $\mathcal{F}^{\mathbb{H}/\mathbb{B}}_d$ and $\mathcal{A}^{\mathbb{H}/\mathbb{B}}_d$.}

\begin{equation}
F^{\mathbb{H}/\mathbb{B}}_d=\cdots + A^{\mathbb{H}/\mathbb{B}}_d\,\log \frac{\delta}{R} + \cdots\, .
\end{equation}
where $\cdots$ stand for other (typically scheme-dependent) terms, while $\delta$ stands for the $UV$ short-distance cut-off, which, on dimensional grounds, must appear through the combination $\delta/ R$ being $R$ the characteristic scale of the space. The relevant observation is that, from the explicit expression of the metrics in \eqref{metrics}, it is evident that a Weyl transformation amounts to a change in $R$. The integrated VEV of the trace of the stress energy tensor immediately follows from $F$ as\footnote{For a Weyl rescaling of the metric by $e^{2\sigma}$, $R$ rescales as $e^{\sigma}$.
Denoting by $\Gamma=\log Z=-F$, one then obtains
 $\int \sqrt{g}\,\langle T^{\mu}_{\mu}\rangle = \frac{\delta \Gamma}{\delta \sigma} = - \frac{\delta F}{\delta \log R} = A_d^{\mathbb{H}/\mathbb{B}}$.}

\begin{equation}
\label{dosdos}
\int_{\mathbb{H}^d/\mathbb{B}^d}\sqrt{g} \, \langle T^{\mu}_{\mu}\rangle = A^{\mathbb{H}/\mathbb{B}}_d\, .
\end{equation}
Thus, the (integrated) trace of the stress-energy tensor corresponds to the coefficient of 
 $\log(\delta/R)$ in the free energy.

In the following we will be interested in the computation of the logarithmic term in the free energy on the spaces  $\mathbb{H}^d$ and $\mathbb{B}^d$, which determines the anomaly. 

On general grounds, such anomalous terms can arise either from bulk anomalies or from boundary anomalies. In even dimensions $d=2n$, for conformally flat spaces such as those of interest in this work, the integrated anomaly is coming solely from the Euler number of the space, since the boundary is an odd sphere which does not admit any anomalous term (other possible boundary terms contain the Weyl tensor and the traceless part of the extrinsic curvature, which vanish for our spaces). This anomaly is given by the simple formula
\begin{equation}
\label{anomaliagen}
\int_{M_{2n}} \sqrt{g} \, \langle T^{\mu}_{\mu}\rangle =-(-1)^{n} \, 2\,a\,\chi \, .
\end{equation}
where $\chi $ is the Euler characteristic of the manifold and $a$ is the $A$-anomaly coefficient. In particular, in four dimensions, with this normalization, a real conformal scalar contributes as $a=1/360$ whereas a Dirac  fermion   contributes as $a=11/360$.

In turn, in odd dimensions $d=2n+1$ the bulk anomaly vanishes and the logarithmic term is coming entirely from the boundary, which in this case is an even-dimensional sphere and thus admits an $A$-type anomaly proportional to its Euler number. In section \ref{centralcharges}, we will analyze these boundary anomalies, leading to the prediction of boundary central charges.

\section{Spaces of even dimension}\label{even}

\subsection{Free fields in $\mathbb{H}^{2n}$}


\subsubsection{Scalars in $\mathbb{H}^{2n}$}

In  this case the $A$-anomaly coefficients have been computed in \cite{Bytsenko:1995ak}. Let us quote the relevant coefficients for completeness

\begin{equation}
\label{scalarsH2n}
A_2^{\mathbb{H}}= \frac{1}{6} \, \quad A_4^{\mathbb{H}}= -\frac{1}{180} \, ,\quad A_6^{\mathbb{H}}=\frac{1}{1512} \, .
\end{equation}

From \eqref{anomaliagen}, using that $\chi(\mathbb{H}^{2n})=1$, it follows that 
\bea
&& d=2:\qquad a=\frac{1}{12}\, ,
\nonumber\\
&& d=4:\qquad a=\frac{1}{360}\, ,
\nonumber\\
&& d=6:\qquad a=\frac{1}{3024}\, ;
\label{anocon}
\eea
which agree with the well known coefficients for conformally coupled scalars.

\subsubsection{Half-spin fields in $\mathbb{H}^{2n}$}

 The heat kernel of the square of the Dirac operator $(\gamma^\mu\nabla_\mu )^2 $ at coincident points for a Dirac fermion on $\mathbb{H}^{d}$ is given by   \cite{Camporesi:1995fb},

\begin{equation}
\label{KfermionsH}
\mathcal{K}_{d}^{\mathbb{H}} = {\bf 1}\  \frac{2^{d-3}\,V_{\mathbb{H}^d}\, \Gamma\big(\frac{d}{2}\big)}{\pi^{\frac{d}{2}+1}}\,\int_0^{\infty}d\lambda\,\mu_d(\lambda) \, e^{-t\lambda^2}\,,
\end{equation}
\begin{equation}
  \mu_d(\lambda) = \frac{\pi}{2^{2d-4}}\,\Big| \frac{\Gamma\big(\frac{d}{2}+i\lambda\big)}{\Gamma\big(\frac{d}{2}\big)
  \,\Gamma\big(\frac{1}{2}+i\lambda\big)}\Big|^2\, ,
\nonumber
\end{equation}
where ${\bf 1}$ is the identity matrix of size $d_s\times d_s$ and $d_s= 2^{d/2}$ represents the dimension of the spinor in even dimensions.
Let us now compute the logarithmic term in the free energy. The contribution to the free energy of the Dirac spinor on $\mathbb{H}^d$ is given by 

\begin{equation}
\label{FfermionsH}
\mathcal{F}^{\mathbb{H}}_{d} = \frac{1}{2} \int_{\delta^2} \frac{dt}{t}\, {\rm Tr}\,  \mathcal{K}_{d}^{\mathbb{H}}\, ,
\end{equation}
%
where $V_{\mathbb{H}^d}$ stands for the regularized volume of the hyperbolic space (see \textit{e.g.} \cite{Rodriguez-Gomez:2017kxf}), while
$\delta$ represents a UV short-distance cutoff.
The trace adds a factor $d_s$.

Starting with the case $d=2$, we have

\begin{equation}
\mathcal{F}^{\mathbb{H}}_2=- \int d\lambda\,\lambda\,\coth(\pi\lambda)\, \int_{\delta^2} \frac{dt}{t}e^{-t\lambda^2}\, .
\end{equation}
The $t$ integral can be done, cutting-off the short time region with $\delta$, finding, in the small-$\delta$ limit

\begin{equation}
\mathcal{F}^{\mathbb{H}}_2=  \int d\lambda\,\lambda\,\coth(\pi\lambda)\, (\gamma_E+\log\delta^2+\log\lambda^2)\, .
\end{equation}
The integral contains power-like divergences which do not contribute to the anomaly.
The only contribution to the logarithmic divergence comes from the term

\begin{equation}
\mathcal{F}^{\mathbb{H}}_2= 2\int d\lambda\,\lambda\,\coth(\pi\lambda)\, \log\delta \, .
\end{equation}
The integral diverges. Renormalizing it by subtraction of the flat space free energy, we have

\begin{equation}
\mathcal{F}^{\mathbb{H}}_2=2 \int d\lambda\,\lambda\,(\coth(\pi\lambda)-1)\, \log\delta =\frac{1}{6} \log\delta \, .
\end{equation}

Let us now consider the four-dimensional  $d=4$ case. A direct application of the formulas above yields

\begin{equation}
\mathcal{F}^{\mathbb{H}}_4=\frac{1}{3} \int d\lambda\,\lambda\,(1+\lambda^2)\coth(\pi\lambda)\, \int_{\delta^2} \frac{dt}{t}e^{-t\lambda^2}\, .
\end{equation}
Integrating over $t$ and keeping the term proportional to $\log\delta$ now yields 

\begin{equation}
\mathcal{F}^{\mathbb{H}}_4=-\frac{2}{3} \int d\lambda\,\lambda\,(1+\lambda^2)\coth(\pi\lambda)\, \log\delta\, .
\end{equation}
Subtracting the infinite contribution of flat space, the logarithmic divergent term comes from the integral, 

\begin{equation}
\mathcal{F}^{\mathbb{H}}_4=-\frac{2}{3} \int d\lambda\,\lambda\,(1+\lambda^2)(\coth(\pi\lambda)-1)\, \log\delta=-\frac{11}{180}\log\delta \, .
\end{equation}

Finally, for $d=6$, repeating the same steps above leads to the integral

\begin{equation}
\mathcal{F}^{\mathbb{H}}_6=\frac{1}{15} \int d\lambda\,\lambda\,(4+5\lambda^2+\lambda^4)(\coth(\pi\lambda)-1)\, \log\delta=\frac{191}{7560}\log\delta \, .
\end{equation}

Compiling our results, we find

\begin{equation}
\label{fermionsH2n}
\mathcal{F}^{\mathbb{H}}_2 =  \frac{1}{6} \log \delta  \,,\qquad \mathcal{F}^{\mathbb{H}}_4=-\frac{11}{180}\log\delta\,,\qquad  \mathcal{F}^{\mathbb{H}}_6= \frac{191}{7560}\log\delta\, .
\end{equation}

From \eqref{anomaliagen}, we now find the following $a$-anomaly coefficients for a Dirac spinor in $d$ dimensions:
\bea
&& d=2:\qquad a=\frac{1}{12}
\nonumber\\
&& d=4:\qquad a=\frac{11}{360}
\nonumber\\
&& d=6:\qquad a=\frac{191}{15120}
\eea
in agreement  with the  known results in the literature (for the $d=6$ case, see {\it e.g.} \cite{Bastianelli:2000hi}).

\subsection{Free fields in $\mathbb{B}^{2n}$}

\subsubsection{Scalars in $\mathbb{B}^{2n}$}

Heat kernel coefficients for Dirichlet and Neumann scalars in the ball $\mathbb{B}^n$ have been computed in 
\cite{levitin} and \cite{Bordag:1995gm}. In particular, from \cite{levitin} one finds ${\rm Tr}\ e^{t\Delta}$ expressed as a power series in $t$, with coefficients $a_{{\rm Dir},k}(d)$
(or $a_{{\rm Neu},k}(d)$)
for the $t^{(k-d)/2}$ term. 
The relevant coefficient that give rise to the logarithmic term  comes from the constant term in this power series expansion. 
Using the formulas in  \cite{levitin} we have

\begin{equation}
F_d^{\mathbb{B}}=\,\frac{1}{(4\pi)^{\frac{d}{2}}}\,a_{{\rm Dir},d}(d)\,V_{\mathbb{S}^{d-1}}\,\log\delta \,,\qquad a_{{\rm Dir},d}(d)=\sum_{j=0}^{d-1}b_{d,j}\,(d-1)^j\, ;
\end{equation}
being the coefficients $b_{k,j}$ those tabulated in \cite{levitin}. Making use of these, we find

\begin{equation}
F_2^{\mathbb{B}}=\frac{1}{6}\log\delta\,,\qquad F_4^{\mathbb{B}}=-\frac{1}{180}\log\delta\,,\qquad F_6^{\mathbb{B}}=\frac{1}{1512}\log\delta\, .
\end{equation}
This leads to
exactly  the same  $A$-anomaly coefficients as in \eqref{anocon} (we use $\chi(\mathbb{B}^{2n})=1$).

In conclusion, we find
\begin{equation}
F_{2n}^{\mathbb{H}}=F_{2n}^{\mathbb{B}}\, .
\end{equation}
The identity is not obvious a priori, given that, in particular,  $\mathbb{H}^{2n}$ is non-compact.


%


\subsubsection{Half-spin fields in $\mathbb{B}^{2n}$}

The heat kernel  for spinors on $\mathbb{B}^{2n}$ has been discussed in 
\cite{Dowker:1995sw}. At coincident points, it admits the expansion 

\begin{equation}
\mathcal{K}(t)=\sum_{k=0}^{\infty} B_{\frac{k}{2}}\,t^{\frac{k}{2}-\frac{d}{2}}\, .
\end{equation}
Thus, in order to extract the logarithmic term in the free energy, the relevant coefficient is that corresponding to $k=d$. 
The coefficients   $B_{k/2}$'s computed in \cite{Dowker:1995sw} include a  factor  $2^n$, corresponding to the dimension of the spinor representation.

Let us begin with Dirac spinors with mixed boundary conditions.
These are implemented by the condition $P_+\psi =0 $ at $r=1$, where $P_+=\frac12 \big(1-i\Gamma^* \Gamma^\mu n_\mu \big)$ in terms of the normal vector $n_\mu$ and a chirality $\Gamma^*$  matrix (see \cite{Dowker:1995sw} for details on the $\Gamma $ matrix notation).
Using the results in section 4 in \cite{Dowker:1995sw} for the $B_{k/2}$'s, we find,

\begin{equation}
\label{fermionsB2n}
\mathcal{F}_2^{\mathbb{B}}=\frac{1}{6}\log\delta\,,\qquad \mathcal{F}_4^{\mathbb{B}}=-\frac{11}{180}\log\delta\,,\qquad \mathcal{F}_6^{\mathbb{B}}=\frac{191}{7560}\log\delta\,.
\end{equation}

Comparing \eqref{fermionsH2n} with \eqref{fermionsB2n} we have that

\begin{equation}
\mathcal{F}_{2n}^{\mathbb{H}}=\,\mathcal{F}_{2n}^{\mathbb{B}}\, .
\end{equation}

Similarly, one can compute the contribution to the logarithmic part of the free energy for fermions with
spectral boundary conditions. They can be implemented by demanding that the negative (positive) modes of the positive (negative) chirality parts of the spinor field vanish at $r=1$ \cite{DEath:1991opx}.
This leads to the following formula for the relevant coefficient in the expansion of the heat kernel \cite{Dowker:1995sw}:
\be
B_{d/2}^{(S)}= 2^{\frac{d}2}\, 2^{-d} (d-1) \left( \frac{F_{d/2}(d)}{\Gamma(\frac{d}2)} +\sqrt{\pi}\  \frac{G_{d/2}(d)}{\Gamma(\frac{d+1}{2})  }\right)\ .
\ee
The   coefficients $ F_{d/2}(d), \ G_{d/2}(d)$  for $d=2,4,6$ are given in eqs. (12), (13) and appendix C of  \cite{Dowker:1995sw}. Strikingly, despite the fact that the general formulas are totally different from the case of mixed boundary conditions,
upon substituting these coefficients, we find exactly the same expressions, \eqref{fermionsB2n}, for  the logarithmic terms. Thus, the conformal anomaly is the same for half-spin fields with spectral and mixed boundary conditions.

\section{Spaces of odd dimension}\label{odd}

\subsection{Free fields in $\mathbb{H}^{2n+1}$}

The free energy in odd dimensional hyperbolic spaces contains an IR  logarithmic divergent term. The origin of this term stems from the fact that the free energy is proportional to the volume of the hyperbolic space $V_{\mathbb{H}^d}$.
From \eqref{muno}, one has 
\be
V_{\mathbb{H}^d}=V_{\mathbb{S}^{d-1}}\int_0^{L} d\rho\, \frac{\rho^{d-1}}{\sqrt{1+\frac{\rho^2}{R^2}}}\ ,
\ee
where $L$ represents the IR cutoff. 
In odd dimensions, the large $L$ expansion contains a logarithmic divergence
(see also appendix A in  \cite{Rodriguez-Gomez:2017kxf}). In particular,  one finds the logarithmic terms
\be
V_{\mathbb{H}^3}=-\frac12 \log \frac{L}{R}\ ,\qquad 
V_{\mathbb{H}^5}=\frac38 \log\frac{L}{R}\ ,\qquad 
V_{\mathbb{H}^5}=-\frac5{16} \log \frac{L}{R}\ .
\ee
There are also power-like divergences that we omitted because they are not relevant for the discussion of the anomaly.
The logarithmic terms give rise to an anomalous contribution under constant rescaling of the metric.
In \cite{Rodriguez-Gomez:2017kxf}, we argued that the origin of this anomaly is an IR/UV relation of the cutoffs
under the conformal map from $\mathbb{H}^{d}$ to $\mathbb{B}^d$, described in the introduction:
\be
L = \frac{R^2}{\delta }\, .
\ee
In this section we will show that, under this identification, the boundary anomalies
on $\mathbb{H}^{2n+1}$ {\it exactly} coincide with the boundary anomalies on $\mathbb{B}^{2n+1}$. This must be the case, since
these spaces are related by a conformal transformation and the boundary anomaly is conformal invariant \cite{Solodukhin:2015eca}.\footnote{In particular, in the case at hand, note that the conformal transformation maps the two boundaries one to the other.
}
Thus the agreement shows that the IR/UV identification leads to the expected result and confirms that free
conformal field theory on the uncompact $\mathbb{H}^{2n+1}$ space can have boundary conformal anomalies.

\subsubsection{Scalars in $\mathbb{H}^{2n+1}$}

The free energy for scalars in $\mathbb{H}^{2n+1}$ was computed in \cite{Rodriguez-Gomez:2017kxf}, extending
previous results to higher dimensional hyperbolic spaces. 
One has  \cite{Rodriguez-Gomez:2017kxf}


\begin{equation}
\label{scalarsHHH}
F_3^{\mathbb{H}}=\frac{1}{48}\log  \frac{L}{R}\,,\qquad F_5^{\mathbb{H}}=-\frac{17}{11520}\log  \frac{L}{R}\,,\qquad F_7^{\mathbb{H}}=\frac{367}{1935360}\log  \frac{L}{R}\, .
\end{equation}

\subsubsection{Half-spin fields in $\mathbb{H}^{2n+1}$}

The free energy for fermions in $\mathbb{H}^{2n+1}$ follows from the same formula  \eqref{FfermionsH}, where the expression for the heat kernel at coincident points is given in \eqref{KfermionsH}. Particularizing this expression to the relevant cases, we find that the $\mu_d(\lambda)$ is a polynomial in $\lambda$, so that the $\lambda$ integral can be easily done. We obtain

\begin{eqnarray}
\label{KfermionH2n+1}
\mathcal{K}_3^{\mathbb{H}}=\frac{1}{16\pi^{\frac{3}{2}}\,t^{\frac{3}{2}}}\,(2+t)\,,\qquad \mathcal{K}_5^{\mathbb{H}}=\frac{1}{384\,\pi^{\frac{5}{2}}\,t^{\frac{5}{2}}}\,(12+20t+9t^2)\,,\nonumber
\\
 \mathcal{K}_7^{\mathbb{H}}=\frac{1}{15360\pi^{\frac{7}{2}}\,t^{\frac{7}{2}}}\,(120+420t+518t^2+225t^3)\, .
\end{eqnarray}
One now notices that $\mathcal{K}_{2n+1}^{\mathbb{H}}$ does not contain any $t$-independent term.
As a result, there is no logarithmic divergence in the free energy $\propto \int \frac{dt}{t}\, \mathcal{K}_d$.  Therefore, one concludes that 

\begin{equation}
\label{fermionH2n+1}
\mathcal{F}_{2n+1}^{\mathbb{H}}=0\,.
\end{equation}

\subsection{Free fields in $\mathbb{B}^{2n+1}$}

\subsubsection{Scalars in $\mathbb{B}^{2n+1}$}


In the even dimensional case, we have considered scalar fields
with Dirichlet boundary conditions and reproduced
known results. Since the odd dimensional cases with $d=5,7$ are new, we will include the case of Neumann or, more general, 
 Robin boundary conditions (also called `mixed' or `generalized Neumann' boundary conditions),
defined by 
\be
\left( \partial_n \phi - \gamma \phi \right)\Big|_{\mathbb{S}^{d-1}}=0\ .
\ee
Conformal invariant boundary conditions are obtained in two cases: Dirichlet or
the conformal invariant Robin condition corresponding to
\be
\left(\partial_n \phi +\frac{(d-2)}{2(d-1)} K \phi \right)\bigg|_{\mathbb{S}^{d-1}}=0\ .
\ee
This is invariant under the conformal transformation that leaves the action invariant,
$g_{\mu\nu}\to e^{2\sigma} g_{\mu\nu}$, $\phi\to \phi\ e^{\frac12 (d-2)\sigma}$.
Since $K(\mathbb{S}^{d-1})=d-1$, we obtain that the conformal Robin condition
corresponds to  
\be
\gamma=1- \frac{d}{2}\ .
\ee

Using the formulas  in \cite{levitin}, we now have

\begin{equation}
F_d^{\mathbb{B}}=\frac{1}{(4\pi)^{\frac{d-1}{2}}}\,a_{{\rm Dir/Neu},d}(d)\,V_{\mathbb{S}^{d-1}}\,\log\delta \, .
\end{equation}
where the Dirichlet and (generalized) Neumann boundary condition
cases are given by

\begin{equation}
    a_{{\rm Dir},d}(d)=\sum_{j=0}^{d-1}b_{d,j}\,(d-1)^j\,, \qquad a_{{\rm Neu},d}(d)=\sum_{j=0}^{d-1}c_{d,j}\,(d-1)^j\, ,
\end{equation}
with the coefficients $b_{k,j},\ c_{k,j}$  computed in
 \cite{levitin}.

For the Dirichlet case,  using the results of \cite{levitin} for 
$b_{d,j},\ c_{d,j}$ , we find

\begin{equation}
F_3^{\mathbb{B}}=-\frac{1}{48}\log\delta\,\qquad F_5^{\mathbb{B}}=\frac{17}{11520}\log \delta\,,\qquad F_7^{\mathbb{B}}=-\frac{367}{1935360}\log \delta\, .
\end{equation}
Under the UV/IR identification $\delta\sim 1/L$. Thus we see that these
expressions exactly reproduce the corresponding expressions 
\eqref{scalarsHHH} for $\mathbb{H}^{2n+1}$.

In turn, for the (generalized) Neumann case, we find

\begin{eqnarray}
&& F_3^{\mathbb{B}}=\Big( \frac{7}{48}+\frac{\gamma}{2}(1+\gamma)\Big)\, \log\delta\,,
\\ \nonumber 
&& F_5^{\mathbb{B}}=\Big(\frac{\gamma^4}{24}+\frac{\gamma^3}{4}+\frac{13\gamma^2}{24}+\frac{\gamma}{2}+\frac{1873}{11520} \Big)\log \delta\,,\\ \nonumber 
&& F_7^{\mathbb{B}}=\Big(\frac{\gamma^6}{720}+\frac{\gamma^5}{48}+\frac{\gamma^4}{8}+\frac{55\gamma^3}{144}+\frac{149\gamma^2}{240}+\frac{\gamma^2}{2}+\frac{291217}{1935360}\Big)\log \delta\, .
\end{eqnarray}
%


%



\subsubsection{Half-spin fields in $\mathbb{B}^{2n+1}$}

Just like in the even-dimensional $\mathbb{B}^{2n}$ case discussed earlier, the relevant coefficients $B_{\frac{d}{2}}$ in the heat kernel expansion for spinors on $\mathbb{B}^{2n+1}$ can be obtained from the 
expressions given in \cite{Dowker:1995sw}. They are given by

\begin{eqnarray}
 \nonumber B_{\frac{3}{2}} &=& -\frac{2^{-d}\,\sqrt{\pi}\,d_s}{64\,\Gamma\big(\frac{d}{2}\big)}\,(d-1)(d-3)\Big|_{d=3}=0\,,\\   B_{\frac{5}{2}} &=& \frac{2^{-d}\,\sqrt{\pi}\,d_s}{122880}\,(d-1)\,(d+1)(d-5)(89d-263)\Big|_{d=5}=0\, ,\\
 \nonumber B_{\frac{7}{2}}
 &=& \frac{2^{-d}\,\sqrt{\pi}\,d_s}{495452160}\,(d-1)(d-7)
\nonumber\\
&& \times (393039+368952d-147742d^2-33848d^3+9167d^4)\Big|_{d=7}=0\,. \nonumber \\ 
\end{eqnarray}
We see that all the relevant Seeley-de Witt coefficients just vanish, thus leaving us with the result

\begin{equation}
\mathcal{F}_{2n+1}^{\mathbb{B}}=0\,.
\end{equation}
Obviously, this agrees with \eqref{fermionH2n+1}, so that it holds that $\mathcal{F}_{2n+1}^{\mathbb{B}}= \mathcal{F}_{2n+1}^{\mathbb{H}}$.

For spin $1/2$ fields with
spectral boundary conditions we find the same result.
Specifically,
\be
B_{d/2}^{(S)}=2^{\frac{d+1}2}\ 2^{-d}  (d-1) \left( \frac{F_{d/2}(d)}{\Gamma(\frac{d+1}2)} +\sqrt{\pi}\  \frac{G_{d/2}(d)}{\Gamma(\frac{d}{2})  }\right) =0
\ee
for $d=3,5,7$. The cancellation is striking, given  the complicated form of the coefficients $F_n(d)$, $G_n(d)$ 
(see \cite{Dowker:1995sw}).

\section{Coefficients of boundary anomalies in $d=3,5,7$ dimensions}\label{centralcharges}

For even-dimensional manifolds without boundary, formulas for the integrated  conformal anomaly  as
a functional of the curvature of the space have been extensively discussed in the literature starting with \cite{Duff:1977ay}, in particular in dimensions 2, 4, 6.
There are two types of anomalies, the $A$-type anomaly given in terms of the Euler characteristic of the space, and
$B$-type anomalies  built from the Weyl tensor and covariant derivatives \cite{Deser:1993yx}. 


When boundaries are present, they can support extra terms in the effective action contributing to the anomaly. The presence of these boundary terms in the anomaly is perhaps more striking in odd-dimensional spacetimes, for which the standard lore says that no conformal anomaly exists.  Even though the complete classification of these boundary terms is not known beyond $d=4$  (see \cite{Herzog:2015ioa,Solodukhin:2015eca,Fursaev:2016inw} for discussions), 
the generic structure of the boundary anomaly terms seems to be as follows: there is an ``$A$-anomaly" contribution arising from the Euler characteristic $\chi$ of the boundary and a ``$B$-anomaly" contribution constructed out of the Weyl tensor and the
traceless part  of the extrinsic curvature
${K}_{ab}$, 
$\hat{K}_{ab}=K_{ab}-\frac{1}{d-1}\gamma_{ab} K$, where $\gamma_{ab}$ is the induced metric on the boundary.
Many examples of such terms are constructed in \cite{Solodukhin:2015eca}. In consequence, for spaces which  are conformally flat --for which the corresponding Weyl tensor vanishes-- and have a boundary with vanishing  $\hat{K}_{ab}$, the putative $B$-type terms of the boundary anomaly would vanish.\footnote{$\hat K_{ab}$ vanishes when the extrinsic curvature is of the form
${K}_{ab}= e^{\sigma(x)} \gamma_{ab}$. In particular,
this occurs in a class of spaces 
where global Gaussian normal coordinates exist such that the metric is of the form
 $ds^2=dr^2+ f(r,x) h_{ab}(x) dx^a dx^b$. Then ${K}_{ab}= \frac12 \partial_r f\big|_{r=r_0} h_{ab}$,
where $r=r_0$ is the location of the boundary.}
Note that the spaces $\mathbb{H}^{2n+1}$ and $\mathbb{B}^{2n+1}$ considered here are in this class.
We conjecture that for any CFT defined
on geometries $M_{2n+1}$ with vanishing Weyl tensor and vanishing
$\hat K_{ab}$, the integrated conformal anomaly is given by the formula:

\begin{equation}
\label{anomalodd}
\int_{M_{2n+1}} \sqrt{g} \, \langle T^{\mu}_{\mu}\rangle = (-1)^{n} a_B\  \chi(\partial M_{2n+1}) \, ,
\end{equation}
where $a_B$ carries a label ``$B$" (Boundary) to distinguish from the usual $A$-anomaly coefficient $a$ multiplying the Euler number of the full space. We stress once again that, for general spaces, the complete formula is expected to contain, in addition,  terms with various combinations of the Weyl tensor and $\hat K_{ab}$ (see \cite{Solodukhin:2015eca}). 

For odd-dimensional spaces the bulk anomalies vanish, and hence the anomalies which we find are coming solely from the boundary as in \eqref{anomalodd}. In three-dimensions, where in fact the full boundary anomaly has been deduced from first principles in \cite{Solodukhin:2015eca,Fursaev:2016inw} (see also \cite{Herzog:2017kkj}), the  coefficient $a_B$ was computed in \cite{Fursaev:2016inw} for conformal scalars with Dirichlet and Robin boundary conditions and fermions with mixed boundary conditions.
In this paper, as an application of our results, we shall compute, for the first time, the contribution to the anomaly coefficient
$a_B$ from conformal scalars and massless fermions in $d=5,7$.
We stress that the result for the coefficients should be universal and apply
to any other spaces (including non-conformally flat spaces), as long as the  \eqref{anomalodd} gives the complete
boundary contribution to the anomaly when the space
is conformally flat and the boundary has    $\hat{K}_{ab}=0$.

On the other hand, for the even-dimensional spaces $\mathbb{H}^{2n}$ and $\mathbb{B}^{2n}$ considered here,  since the boundaries are $2n-1$ dimensional spheres, one has $\chi(\mathbb{S}^{2n-1})=0$
and there is no contribution to the anomaly proportional to $\chi(\partial M_{d})$.
 In this $d$ even case, the anomalies computed here are 
 $A$-anomalies \eqref{anomaliagen} associated with the Euler number of the space
 $\chi(M_{d})$.\footnote{
 Note that, for  even-dimensional balls, the bulk densities vanish, which implies that  the conformal anomaly  is all boundary. We thank the anonymous referee for making this point.}
 Consistently, our results recover the expected central charges for scalars and spinors in various dimensions.


\subsubsection*{$d=3$}

Consider (\ref{anomalodd}) applied to the Euclidean ball $\mathbb{B}^{3}$:
\begin{equation}
\label{anomalB}
\int_{\mathbb{B}^{3}} \sqrt{g} \, \langle T^{\mu}_{\mu}\rangle = - a_B\, \chi(\mathbb{S}^{2}) = -2 a_B\, .
\end{equation}
From \eqref{dosdos} and using the results of section 4, we obtain that conformal scalars with Dirichlet boundary conditions and massless half-spin fields with mixed or spectral boundary conditions
contribute to the anomaly by  the following coefficients:
\bea
&& {\rm Dirichlet\ scalar}:\qquad \qquad\qquad \ \   a_{B} = \frac{1}{96}\  
\nonumber\\
&& {\rm Spin\ 1/2\ (mixed\ or\ spectral)}:\quad a_{B} = 0\ .
\eea

Similarly, we can compute the boundary anomaly coefficient for  a scalar field with conformal Robin boundary conditions.

From the results of section 4, we  obtain

\be
{\rm Robin\ scalar}:\quad a_{B} = -\frac{7}{96}-\frac{1}{4}\gamma (\gamma+1)\ .
\ee
$\gamma =0 $ corresponds to Neumann boundary conditions. 
For conformal Robin boundary conditions, $\gamma=1-d/2=-1/2$
hence  $a_{B} =-1/96$. 

These results are in agreement with \cite{Fursaev:2016inw}.

\subsubsection*{$d=5$}

Considering (\ref{anomalodd}) applied to the ball $\mathbb{B}^{5}$, we now have
\begin{equation}
\label{anomalBB}
\int_{\mathbb{B}^{5}} \sqrt{g} \, \langle T^{\mu}_{\mu}\rangle = a_B \,\chi(\mathbb{S}^{4}) =2 a_B\, .
\end{equation}
Comparing with \eqref{dosdos} and using the results of section 4,  we now find   the following boundary anomaly coefficients:
\bea
&&{\rm Dirichlet\ scalar}:\quad a_{B} = \frac{17}{23040}
\nonumber\\
&& {\rm Spin\ 1/2\ (mixed\ or\ spectral)}:\quad a_{B} = 0\ .
\eea
and
\be 
{\rm Robin\ scalar}:\quad a_{B} =\frac{\gamma ^4}{48}+\frac{\gamma ^3}{8}+\frac{13 \gamma ^2}{48}+\frac{\gamma
   }{4}+\frac{1873}{23040}\ .
\ee
For conformal Robin boundary conditions, $\gamma =-3/2$ and $a_{B} = -\frac{17}{23040}$.

\subsubsection*{$d=7$}

Similarly, we now have
\begin{equation}
\label{anomalBBB}
\int_{\mathbb{B}^{7}} \sqrt{g} \, \langle T^{\mu}_{\mu}\rangle = -a_B  \chi(\mathbb{S}^{6}) =-2 a_B\, .
\end{equation}
with
\bea
&&{\rm Dirichlet\ scalar}:\quad a_{B} = \frac{367}{3870720}\  ,
\nonumber\\
&& {\rm Spin\ 1/2\ (mixed\ or\ spectral)}:\quad a_{B} = 0\ .
\eea
and
\be 
{\rm Robin\ scalar}:\quad a_{B} =-\Big(\frac{\gamma ^6}{1440}+\frac{\gamma ^5}{96}+\frac{\gamma ^4}{16}+\frac{55 \gamma
   ^3}{288}+\frac{149 \gamma ^2}{480}+\frac{\gamma }{4}+\frac{291217}{3870720}\Big)\ .
\ee
For conformal Robin boundary conditions, $\gamma =-5/2$ and $a_{B} = - \frac{367}{3870720}$.

%

Note that, in all cases, conformal Robin boundary conditions give the same value for the $a_B$-central charge as the Dirichlet scalar, but with opposite sign.

\section*{Acknowledgements}

We thank an anonymous referee for pointing out an incorrect statement
in an earlier version of this paper.
D.R-G is partly supported by the Ramon y Cajal grant RYC-2011-07593 as well as the EU CIG grant UE-14-GT5LD2013-618459, the Asturias Government grant FC-15-GRUPIN14-108 and Spanish Government grant MINECO-16-FPA2015-63667-P. J.G.R. acknowledges financial support from a MINECO grant FPA2016-76005-C2-1-P
and  MDM-2014-0369 of ICCUB (Unidad de Excelencia `Mar\'ia de Maeztu').


\begin{thebibliography}{99}


\bibitem{Dowker:1989ue}
  J.~S.~Dowker and J.~P.~Schofield,
  ``Conformal Transformations and the Effective Action in the Presence of Boundaries,''
  J.\ Math.\ Phys.\  {\bf 31} (1990) 808.
  
  
\bibitem{Fursaev:2013mxa} 
  D.~V.~Fursaev,
  ``Quantum Entanglement on Boundaries,''
  JHEP {\bf 1307}, 119 (2013)
  [arXiv:1305.2334 [hep-th]].
        
\bibitem{Jensen:2013lxa} 
  K.~Jensen and A.~O'Bannon,
  ``Holography, Entanglement Entropy, and Conformal Field Theories with Boundaries or Defects,''
  Phys.\ Rev.\ D {\bf 88}, no. 10, 106006 (2013)
  [arXiv:1309.4523 [hep-th]].
       
\bibitem{Jensen:2015swa} 
  K.~Jensen and A.~O'Bannon,
  ``Constraint on Defect and Boundary Renormalization Group Flows,''
  Phys.\ Rev.\ Lett.\  {\bf 116}, no. 9, 091601 (2016)
  [arXiv:1509.02160 [hep-th]].
  
\bibitem{Herzog:2015ioa}
  C.~P.~Herzog, K.~W.~Huang and K.~Jensen,
  ``Universal Entanglement and Boundary Geometry in Conformal Field Theory,''
  JHEP {\bf 1601} (2016) 162
  [arXiv:1510.00021 [hep-th]].
   
\bibitem{Fursaev:2015wpa} 
  D.~Fursaev,
  ``Conformal anomalies of CFT's with boundaries,''
  JHEP {\bf 1512}, 112 (2015)
  [arXiv:1510.01427 [hep-th]].
  
\bibitem{Solodukhin:2015eca}
  S.~N.~Solodukhin,
  ``Boundary terms of conformal anomaly,''
  Phys.\ Lett.\ B {\bf 752} (2016) 131
  [arXiv:1510.04566 [hep-th]].
  
\bibitem{Fursaev:2016inw}
  D.~V.~Fursaev and S.~N.~Solodukhin,
  ``Anomalies, entropy and boundaries,''
  Phys.\ Rev.\ D {\bf 93} (2016) no.8,  084021
  [arXiv:1601.06418 [hep-th]].
 
\bibitem{Huang:2016rol} 
  K.~W.~Huang,
  ``Boundary Anomalies and Correlation Functions,''
  JHEP {\bf 1608}, 013 (2016)
  [arXiv:1604.02138 [hep-th]].
  
\bibitem{Herzog:2017xha} 
  C.~P.~Herzog and K.~W.~Huang,
  ``Boundary Conformal Field Theory and a Boundary Central Charge,''
  arXiv:1707.06224 [hep-th].
  
\bibitem{Rodriguez-Gomez:2017kxf} 
  D.~Rodriguez-Gomez and J.~G.~Russo,
  ``Free energy and boundary anomalies on $\mathbb{S}^a\times \mathbb{H}^b$ spaces,''
  JHEP {\bf 1710}, 084 (2017)
  [arXiv:1708.00305 [hep-th]].
  
\bibitem{Herzog:2017kkj} 
  C.~Herzog, K.~W.~Huang and K.~Jensen,
  ``Displacement Operators and Constraints on Boundary Central Charges,''
  arXiv:1709.07431 [hep-th].
  
\bibitem{Bytsenko:1995ak} 
  A.~A.~Bytsenko, E.~Elizalde and S.~D.~Odintsov,
  ``The Conformal anomaly in N-dimensional spaces having a hyperbolic spatial section,''
  J.\ Math.\ Phys.\  {\bf 36}, 5084 (1995)
  [gr-qc/9505047].
  
\bibitem{Camporesi:1995fb} 
  R.~Camporesi and A.~Higuchi,
  ``On the Eigen functions of the Dirac operator on spheres and real hyperbolic spaces,''
  J.\ Geom.\ Phys.\  {\bf 20}, 1 (1996)
  [gr-qc/9505009].
  
\bibitem{Bastianelli:2000hi} 
  F.~Bastianelli, S.~Frolov and A.~A.~Tseytlin,
  ``Conformal anomaly of (2,0) tensor multiplet in six-dimensions and AdS / CFT correspondence,''
  JHEP {\bf 0002}, 013 (2000)
  [hep-th/0001041].
  
  \bibitem{levitin} M.~Levitin,  ``Dirichlet and Neumann heat invariants for Euclidean balls",
  Diff. Geom. and Appl. 8 (1998) 35.
  
\bibitem{Bordag:1995gm} 
  M.~Bordag, E.~Elizalde and K.~Kirsten,
  ``Heat kernel coefficients of the Laplace operator on the D-dimensional ball,''
  J.\ Math.\ Phys.\  {\bf 37}, 895 (1996)
  [hep-th/9503023].
  
\bibitem{Dowker:1995sw} 
  J.~S.~Dowker, J.~S.~Apps, K.~Kirsten and M.~Bordag,
  ``Spectral invariants for the Dirac equation on the d ball with various boundary conditions,''
  Class.\ Quant.\ Grav.\  {\bf 13}, 2911 (1996)
  [hep-th/9511060].
  
\bibitem{DEath:1991opx}
  P.~D.~D'Eath and G.~Esposito,
  ``Spectral boundary conditions in one loop quantum cosmology,''
  Phys.\ Rev.\ D {\bf 44} (1991) 1713
  [gr-qc/9507005].


\bibitem{Duff:1977ay} 
  M.~J.~Duff,
  ``Observations on Conformal Anomalies,''
  Nucl.\ Phys.\ B {\bf 125}, 334 (1977).
  
 
  
\bibitem{Deser:1993yx} 
  S.~Deser and A.~Schwimmer,
  ``Geometric classification of conformal anomalies in arbitrary dimensions,''
  Phys.\ Lett.\ B {\bf 309}, 279 (1993)
  [hep-th/9302047].
  
  
 
  

  
 
 

  
   

  

  
\end{thebibliography}
\end{document}